# Magnetism near Half-filling of a van Hove Singularity in Twisted Graphene Bilayer

Yi-Wen Liu, Jia-Bin Qiao, Chao Yan, Yu Zhang, Si-Yu Li, and Lin He*

Center for Advanced Quantum Studies, Department of Physics, Beijing Normal University, Beijing, 100875, People's Republic of China

Correspondence and requests for materials should be addressed to L.H. (e-mail: helin@bnu.edu.cn).

**Twisted graphene bilayers (TGBs) have low-energy van Hove singularities (VHSs) that are strongly localized around AA-stacked regions of the moiré pattern. Therefore, they exhibit novel many-body electronic states, such as Mott-like insulator and unconventional superconductivity. Unfortunately, these strongly correlated states were only observed in magic angle TGBs with the twist angle $\theta \approx 1.1°$, requiring a precisely tuned structure. Is it possible to realize exotic quantum phases in the TGBs not limited at the magic angle? Here we studied electronic properties of a TGB with $\theta \sim 1.64°$ and demonstrated that a VHS splits into two spin-polarized states flanking the Fermi energy when the VHS is close to the Fermi level. Such a result indicates that localized magnetic moments emerge in the AA-stacked regions of the TGB. Since the low-energy VHSs are quite easy to be reached in slightly TGBs, our result therefore provides a facile direction to realize novel quantum phases in graphene system.**

Twisted graphene bilayers (TGBs) exhibit different electronic properties depending sensitively on the twisted angle[1-8]. This is especially the case at twist angles $\theta < 2°$, for which the Fermi velocity of the TGBs is strongly suppressed and the local density of states of the van Hove singularities (VHSs) becomes dominated by quasi-localized states in the AA-stacked regions of the moiré pattern[9-14]. In such a regime, Coulomb interactions greatly exceed the kinetic energy of the electrons[7,8,15,16]. Therefore, the TGBs with $\theta < 2°$ are expected to exhibit strongly correlated quantum phases that are almost impossible to realize in graphene monolayer. Very recently, a Mott-like insulating phase[7] and unconventional superconductivity[8] are observed in magic angle TGBs with $\theta \approx 1.1°$, where the low-energy Fermi velocity of the system is almost zero. These results have attracted much attention both in experiment and theory. However, obtaining the TGB with exactly the magic angle is extremely difficult, preventing the experimental exploration of the strongly correlated quantum phases.

In this paper, we showed experimentally that it is possible to realize novel correlated quantum phases in slightly TGBs not limited at the magic angle. Our experiment demonstrated that a VHS with divergent density-of-state (DOS) splits into two spin-polarized peaks when tuning the Fermi level to the VHS of a 1.64 °TGB, indicating that localized magnetic moments emerge in the AA-stacked regions of the TGB. Our experiment further demonstrated that the localized magnetic moments can be easily switched on and off by changing the occupation of the VHS.

In our experiment, large-area aligned graphene monolayer was grown on copper foils[17] and then we fabricated large-scale TGB with a uniform twist angle based on the aligned graphene, as schematically shown in Fig. 1a (see Methods and Supplementary Fig. 1 for the growth and Supplementary Fig. 2 for further characterization of aligned graphene monolayer[18]). The rotation angle $\theta$ can be controlled manually according to the edges of the graphene flakes. The obtained TGB structures with controlled $\theta$ were transferred from the Cu foils onto single-crystal $SrTiO_3$ substrates, which have been annealed in vacuum to obtain large-area terraces advanced, as shown in Fig. 1b as an example. Our scanning tunneling microscope (STM) and spectroscopy (STS) measurements were carried out in the TGBs with controlled twist angle on $SrTiO_3$.

Figures 1c-1e show representative STM images of three different TGBs with different periods $D$ of moiré pattern. The obtained TGBs are quite uniform and exhibit identical period of the moiré pattern on different terraces of the SrTiO$_3$. The twist angles $\theta$ can be estimated by $D = a/[2\sin(\theta/2)]$ with $a = 0.246$ nm the lattice constant of graphene[2-6]. Obviously, our experiment provides a general strategy to study twist engineering in graphene bilayer. Such a method could also extend to other two-dimensional systems.

The TGBs exhibit two low-energy VHSs (see Supplementary Fig. 3 for STS spectra of different TGBs[18]), which depend sensitively on the twist angle. With decreasing the twist angle of the TGBs, the VHSs are approaching the Fermi level and the Fermi velocity decreases dramatically[4,9-13]. To explore possible exotic quantum phases, we systematically studied electronic properties of a TGB with $\theta \approx 1.64° \pm 0.10°$, where the kinetic energy of the low-energy quasiparticles is expected to be strongly suppressed. Figure 2a shows a typical STM image of the TGB. The period of the moiré pattern is about 8.6 ±0.5 nm and the bright dots in the STM image correspond to the AA-stacked regions. The twisting between the two adjacent layers leads to a relative shift of the two Dirac cones $|\Delta K| = 2|K|\sin(\theta/2)$ in reciprocal space, where $K$ is the reciprocal-lattice vector. To the zeroth order, a finite interlayer coupling results in two saddle points appearing at the intersections of the two Dirac cones and leads to the strongly suppressing of the Fermi velocity, as shown in Fig. 2b. Theoretically, the Fermi velocity of the $\theta \approx 1.64°$ TGB is reduced to $0.15 \times 10^6$ m/s, which is much smaller than that of graphene monolayer $\sim 1.0 \times 10^6$ m/s. It means that the kinetic energy of the low-energy quasiparticles in the $\theta \approx 1.64°$ TGB is reduced to about 2% of that in graphene monolayer. Figure 2c shows three representative STS spectra recorded at 1.7 K at different positions of the moiré pattern (the energy resolution of the spectra is about 0.5 meV[18]). In slightly TGBs around the magic angle, the low-energy VHSs usually exhibit features beyond the description of the tight-binding model[7,16]. Low energy pronounced VHS peaks, which may arise from complex moiré bands in slightly TGBs[9], are clearly observed in the tunneling spectra. A notable feature of these low-energy electronic states is that they are mainly localized in the AA-stacked regions of the moiré pattern. Such a result is shown more clearly in the STS map of Fig. 2d: a pronounced VHS with the energy around the Fermi level is strongly localized in the AA-stacked region. Similar result has been confirmed in all the studied moiré pattern and our experiment demonstrated that there are quasi-localized states periodically in the AA-stacked regions of the TGB.

Theoretically, the quasi-localized state around the Fermi level is expected to exhibit local magnetic moment because that double occupation of this state by two electrons with opposite spins is energetically unfavorable due to the electrostatic Coulomb repulsion $U$[19-26]. The electrostatic Coulomb repulsion will lead to spin splitting of the quasi-localized state and result in two spin-polarized peaks that are symmetric around the Fermi energy $E_F$. To explore the magnetic properties of the system, we carried out STS measurements with the energy resolution of 0.1 meV at temperature 0.4 K (see Supplemental materials for details[18]). From now on, we will focus on the VHS around the Fermi energy, within several millielectron volt. Figure 3a shows a STS spectrum, measured in the AA-stacked regions, that is representative of our findings. The spectrum has two DOS peaks, flanking the Fermi energy, separated in energy by a splitting of ~3.0 meV. We can remove the contribution from other VHSs because that the energy separation between the $E_F$ and the other DOS peaks in the spectra is at least larger than 10 meV, as shown in Fig. 2c. The two well-defined peaks are attributed to the two spin-polarized states that are fully separated by the on-site Coulomb repulsion, which can be estimated to be $e^2/(4\pi\varepsilon d)$. Here $e$ is the electron charge, $\varepsilon$ is the effective dielectric constant including screening and $d$ is the effective linear dimension of each site, which should be the same length scale as the period of the moiré pattern[7]. Previously, similar spin splitting has been reported in the localized state induced by hydrogen atoms absorbed on graphene[26]. In our experiment, the spatial distribution of the VHS, i.e., the period of the moiré pattern, is about 8.6 nm, which is much larger than that, ~2 nm, of the localized state induced by hydrogen atoms absorbed on graphene[26]. Therefore, the observed splitting, ~3.0 meV, in our experiment is much smaller than that, ~20 meV, observed on the absorbed hydrogen atoms[26].

According to the theory originally proposed by Anderson[19], the magnetism of the quasi-localized state should be tuned by changing the occupation of the split states. We realized the transition from a magnetic state to a nonmagnetic state, or vice versa, by tuning the energy position of the quasi-localized state with respect to the Fermi level. In our experiment, there is a slight spatial variation of charge transfer between the STO and the TGB, and the Fermi level at different moiré of the TGB on the STO terraces could varies 1.0-1.5 meV. A slight variation of the Fermi level of only 1.0-1.5 meV could switch on or off the local magnetic moment, which provides unprecedented opportunity to tune the magnetic properties of the TGBs. Figure 3b and 3c show two representative spectra that both the states with different spins are completely unoccupied and occupied, respectively. For both the cases, the spin splitting vanishes and only one peak with degenerate spin states can be observed. Here we should point out that the features of other VHSs away from the Fermi energy are not affected by the slight variation of the doping. Figure 3d summarized the spin splitting of the VHS around the Fermi level as a function of graphene doping observed in our experiment. Our experimental result can be described quite well by the calculation in terms of the Anderson impurity model[19,26]. By taking into account three important parameters: 1) the energy position of the localized state $E_d$ with respect to the Fermi level, 2) the energy width of the localized state $\Delta$, and 3) the Coulomb repulsion in the localized state $U$, this model can describe the conditions for the absence or existence of the localized magnetic moment in the AA-stacked regions (see Supplemental materials for details of the calculation[18]). For a localized state with fixed $\Delta$ and $U$, it is facile to switch on or off the local magnetic moment of the localized state by changing its occupation, as also shown in Fig. 3d. In the calculation, $2\Delta = 1.7$ meV and $U = 3.3$ meV are used. The Coulomb repulsion $U$ for an isolated moiré pattern roughly estimated according to $e^2/(4\pi\varepsilon d)$ is about 17 meV if we assume $\varepsilon = 10\varepsilon_0$ and $d = 8.6$ nm. In our experiment, the electrons in the surrounding moiré patterns of the TGB may reduce the on-site Coulomb repulsion and result in the relatively small spin splitting.

The result that the two split peaks are the spin-polarized states is further confirmed by carrying out measurements in magnetic fields. Figure 4a shows STS spectra of the two peaks in different magnetic fields. The doping of the two states various slightly with the magnetic fields because of the redistribution of charges between the graphene sheets and the supporting substrate in the presence of magnetic fields, as observed previously[27,28]. A notable feature is that the energy separations of the two peaks increase linearly with the magnetic fields (Fig. 4b), which further demonstrated that the two peaks are two spin-polarized states. Similar result has also been observed at other temperatures (see Supplementary Fig. 4 for STS spectra recorded at 1.7 K[18]). To quantitatively describe the linear relationship, we define the effective gyromagnetic ratio as $g_{eff} = \mu_B^{-1}(\partial \Delta E/\partial B)$, and the linear portion yields a $g_{eff} = 17$ at 0.4 K and $g_{eff} = 12$ at 1.7 K. The observed energy separations in magnetic fields are much larger than that of the Zeeman splitting (the effective gyromagnetic ratio for the Zeeman splitting is usually about 2) (Fig. 4b). Such a result is quite reasonable because the spin splitting in this work arises from the Coulomb interactions. The increase of the energy separations with magnetic fields should be mainly attributed to the enhanced on-site Coulomb interactions in magnetic fields. The magnetic fields can generate Landau gaps (gaps between Landau levels) in graphene[27,28] and strongly change spatial distributions of the electrons in graphene, which are expected to affect the effective dielectric constant and the effective linear dimension of each site *d*. Therefore, the spin splitting of the VHS is strongly enhanced by the magnetic fields. The above experimental results have been verified in different moiré patterns of the 1.64 °TGB. It indicates that there are magnetic moments localized periodically in the AA-stacked regions when the VHS in the TGB is half filled.

For the Mott-like insulating phase observed in the magic-angle TGB[7], the VHS also splits into two peaks at half filling, which is similar as that observed in the 1.64 °TGB in this work. However, there are three important differences between them. First, the observed splitting ~ 3 meV in this work is much larger than the Mott-like insulating gap ~ 0.3 meV. Second, the Mott-like insulating phase is expected to be observed at exactly half filling of the VHS. However, the spin splitting of the VHS is observed even when the VHS is partially filled. Third, the observed splitting in this work increases with the magnetic fields, whereas, the measured Mott-like insulating gap decreases with the magnetic fields[7]. These differences arise from the distinct magnetic nature between the spin splitting of the VHS reported in this work and the Mott-like insulating phase observed in the magic-angle TGB[7].

Interesting magnetic properties of two-dimensional systems around the VHSs were predicted long before graphene was isolated[29,30]. Very recently, the magnetic properties of the TGBs with twist angles $\theta < 2°$ were also studied in theory[31-33]. In the theoretical work[32], the authors considered the Hubbard model in a 1.5 °TGB and predicted that the magnetic moments are mainly localized in the AA-stacked regions, which agrees quite well with our experimental result. Depending on the electrical bias between adjacent layers of the TGB, the authors predicted two possible magnetic orders: one is that the magnetic moments in the adjacent AA regions are parallel; the other is spiral magnetic order where there is a relative 120 ° misalignment between the magnetic moments of neighboring AA regions due to a frustrated antiferromagnetic exchange in the triangular lattice[32]. Further experiment, possibly with the help of spin-polarized STM, should be carried out to detect the exact magnetic coupling between the neighboring AA regions and its dependence on the electrical bias between layers in the slightly TGBs.

In summary, our findings demonstrated that there are magnetic moments localized in AA-stacked regions of the slightly TGBs. This opens the way towards the realization of exotic correlated quantum phases in slightly TGBs not limited at the magic angle. Moreover, our result indicated that the slightly TGB could be an ideal system to study frustrated magnetism in two dimensions because of the triangular lattice of the AA regions.

**Acknowledgements**

We thank Haiwen Liu, Jinhua Gao, and Hua Jiang for helpful discussion. This work was supported by the National Natural Science Foundation of China (Grant Nos. 11674029, 11422430, 11374035), the National Basic Research Program of China (Grants Nos. 2014CB920903, 2013CBA01603). L.H. also acknowledges support from the National Program for Support of Top-notch Young Professionals, support from "the Fundamental Research Funds for the Central Universities", and support from "Chang Jiang Scholars Program".

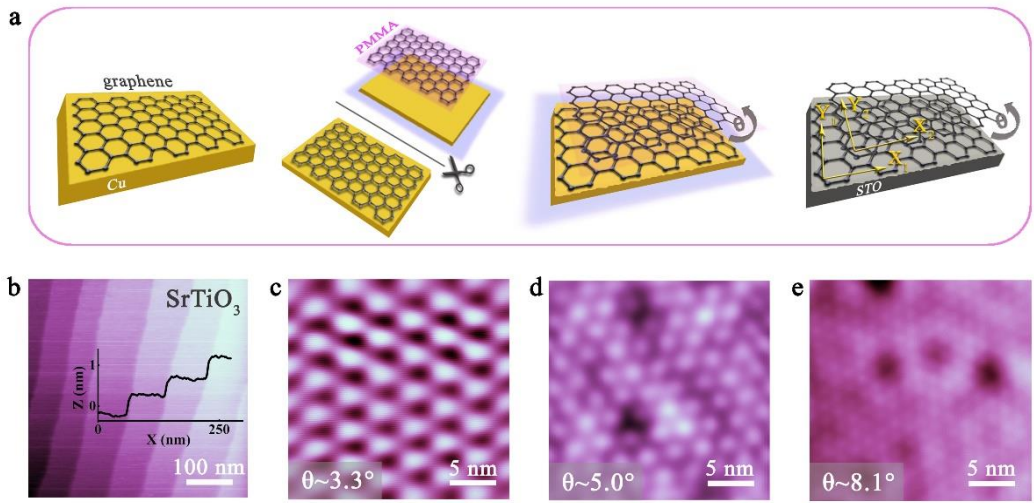

**FIG. 1. Fabrication and characterization of the TGBs.** (a) Illustration showing the fabrication of TGBs with controlled twist angle. Alighed graphene monolayer was grown on a copper foil. Then the graphene monolayer along with the copper foil was cut into two flakes. PMMA was spin-coated on one of the flake and the graphene sheet was transferred onto the other one after that the copper foil was etched by ammonium persulfate. Finally, the TGB with controlled twist angle was transferred onto the 0.7% Nb-doped $SrTiO_3$ (001) substrate. The PMMA was removed by high temperature annealing before STM measurements. (b) STM topographic image of the $SrTiO_3$(STO) surface. The inset is a profile line showing atomic terraces of the STO. (c)-(e) STM images of the TGBs with different twist angles on the STO: (c) $\theta \sim (3.3 \pm 0.4)°$, $V_{sample}$ = -0.7 V, $I$ = 0.2 nA; (d) $\theta \sim (5.0 \pm 0.2)°$, $V_{sample}$ = 0.5 V, $I$ = 0.2 nA; (e) $\theta \sim (8.1 \pm 0.3)°$, $V_{sample}$ = 0.9 V, $I$ = 0.5 nA.

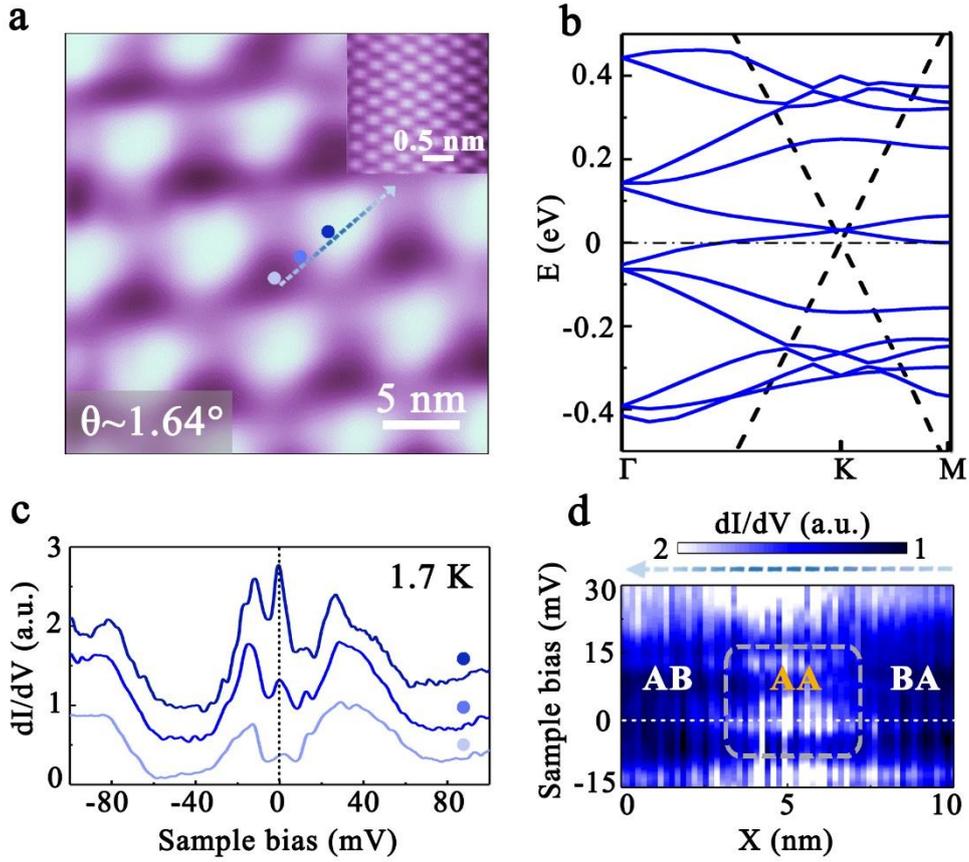

**FIG. 2. STM and STS measurements of a 1.64 ° TGB.** (a) STM topographic image ($V_{sample}$ = 60 mV, $I$ = 0.3 nA) of a TBG with the twist angle $\theta \sim (1.64 \pm 0.10)$ ° and the period $D \sim (8.60 \pm 0.53)$ nm. Inset: atomic-resolution STM image in the AA region of the TGB. (b) Theoretical calculated band structure of the TBG with twist angle $\theta \sim 1.6$ ° (solid blue curves) and that of graphene monolayer with linear band dispersion (dashed black lines). (c) Typical STS spectra of the TBG recorded at three blue dots in panel (a) at temperature 1.7 K. (d) Spatial-resolved dI/dV map obtained along the blue dashed arrow in panel (a).

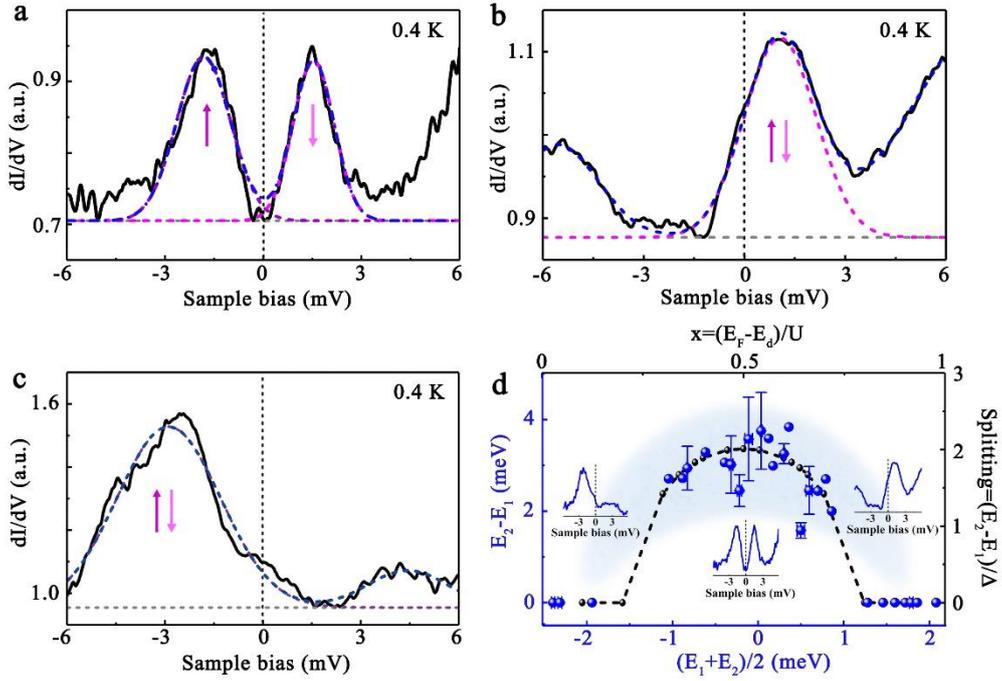

**FIG.3. Spin-split states of a VHS around the Fermi level.** (a)-(c) High-resolution STS spectra of the VHS at temperature 0.4 K. The VHS splits into two spin-polarized peaks, one below and one above the Fermi level, when the VHS is at half filling (a). The spin splitting vanishes when the VHS is fully unoccupied (b) or completely occupied (c). The dashed lines are Gaussian fittings for the VHS. (d) The blue dots are the experiment data of the energy differences between the two spin-polarized peaks near Fermi level as a function of $(E_1+E_2)/2$. $E_1$ and $E_2$ are defined as the energy of two states with opposite spins. The black dashed curve is theoretical spin split based on Anderson impurity model as a function of electron doping. In the calculation, we assumed $\pi\Delta/U = 0.8$.

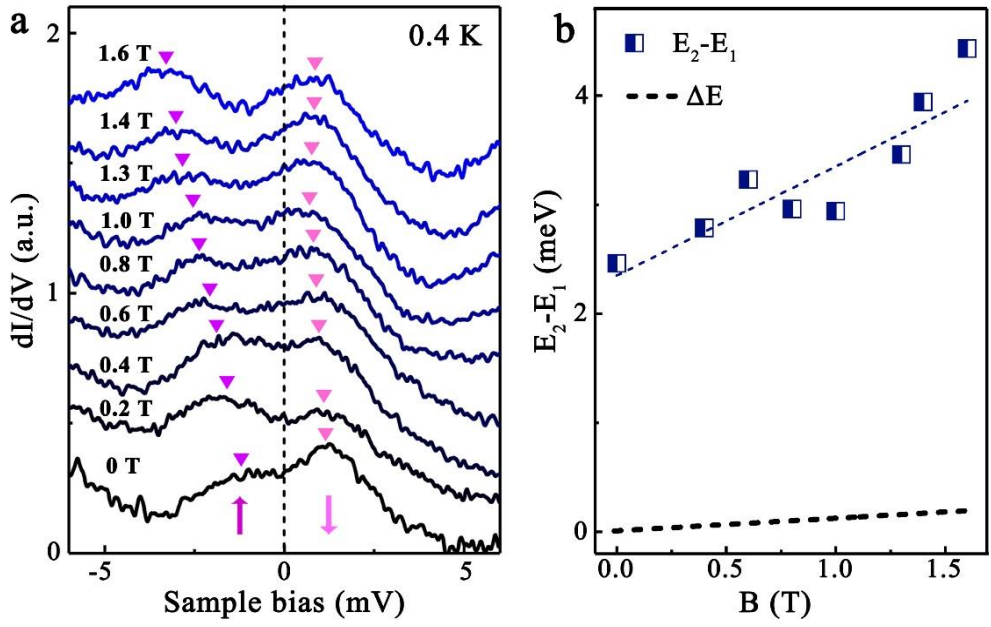

**FIG. 4. The spin-split states in different magnetic fields.** (a) High-resolution STS spectra measured in different magnetic fields and at around 0.4 K. The purple and pink arrows indicate the spin-up and spin-down peaks near the Fermi level around one AA-stacked region of the TGB. Similar result has been confirmed in different AA regions in our experiment. (b) Energy separations of the two peaks as a function of magnetic fields. The black dashed line is the expected Zeeman splitting, $\Delta E = g\mu_B B$ with $g = 2$.